\documentclass[aps,pre,twocolumn,floats]{revtex4}
\usepackage{epsfig}
\usepackage{bm}

\begin{document}

\newcommand{\hide}[1]{}
\newcommand{\tbox}[1]{\mbox{\tiny #1}}
\newcommand{\half}{\mbox{\small $\frac{1}{2}$}}
\newcommand{\const}{\mbox{const}}
\newcommand{\ointt}{\int\!\!\!\!\int\!\!\!\!\!\circ\ }
\newcommand{\intt}{\int\!\!\!\!\int }
\newcommand{\ar}{\mathsf r}
\newcommand{\im}{\mbox{Im}}
\newcommand{\re}{\mbox{Re}}
\newcommand{\sinc}{\mbox{sinc}}
\newcommand{\trc}{\mbox{trace}}
\newcommand{\eexp}{\mbox{e}^}
\newcommand{\bra}{\left\langle}
\newcommand{\ket}{\right\rangle}
\newcommand{\Cn}[1]{\begin{center}{#1}\end{center}}


\title{
Quantum pumping and dissipation: from closed to open systems
}

\author{Doron Cohen}

\affiliation{
Department of Physics, Ben-Gurion University, Beer-Sheva 84105, Israel
}


\begin{abstract}
Current can be pumped through a closed system
by changing parameters (or fields) in time.
The Kubo formula allows to distinguish between dissipative
and non-dissipative contributions to the current.
We  obtain a Green function expression
and an $S$ matrix formula for the associated terms
in the generalized conductance matrix:
the ``geometric magnetism" term 
that corresponds to adiabatic transport;
and the ``Fermi golden rule" term which is responsible
to the irreversible absorption of energy.
We explain the subtle limit of an infinite system,
and demonstrate the consistency with
the formulas by Landauer and B\"{u}ttiker, Pr\'{e}tre and Thomas.
We also discuss the generalization of the fluctuation-dissipation 
relation, and the implications of the Onsager reciprocity. 
\end{abstract}

\maketitle


Linear response theory (LRT) is the traditional theoretical tool
for dealing with the response of driven systems \cite{landau,imry,datta,imryK}.
It offers an expression (the Kubo formula) for the
generalized susceptibility, and hence for the generalized
conductance matrix. It has been realized that in the adiabatic limit
the Kubo formula reduces to an expression
for ``geometric magnetism" \cite{robbins}.
In case of electrical current calculation the latter
gives the ``adiabatic transport" of charge \cite{thouless,avron}.
Outside of the adiabatic regime the response
includes an additional ``dissipation"  term \cite{pmc}.
The latter determines the rate of irreversible energy absorption,
which is caused by Fermi-golden-rule transitions between
energy levels.

Recently there was much interest in analyzing 
the response of {\em open systems} that are connected 
to reservoirs. The analysis has been based on 
the $S$-matrix formalism, leading to the
Landauer formula \cite{imry,datta},  and more generally to the
B\"{u}ttiker, Pr\'{e}tre and Thomas (BPT) formula \cite{BPT}.
A major motivation for the present work is
the realization that the relation between the BPT formula
and the Kubo formula has not been clarified.
In particular the notion of ``adiabatic pumping" in the
context of an open system has been left
obscured, and some confusion has arose regarding
the role of dissipation in the pumping process \cite{SAA,pmp,pmc}.

The purpose of the present work is to analyze 
the response of {\em closed isolated systems} \cite{pmp,pmpMB},  
and in particular to consider the special limit 
of an infinite system (no reservoirs!). 
Thus we are going to construct a bridge between 
the LRT formulation and the BPT formula.  
This is of great practical importance,
because the assumed open geometry of the $S$-matrix
formulation is in many cases an idealization.
It is clear that Kubo formula allows a straightforward
incorporation of finite-size, external noise,
environmental and possibly also many-body effects.
A major step in constructing this bridge,
had been taken in Ref.\cite{fisher}, where the authors
start with the Kubo formula for the electrical
conductivity and end up with the Landauer formula
which relates the conductance to the transmission
of the device.
We are going to see that the general case,
which deals with the {\em generalized} conductance
matrix, and hence incorporates adiabatic transport,
is much more subtle.


Consider a {\em closed isolated system} whose Hamiltonian ${\cal H}$
depends on several control parameters $x_j$.
An example is presented in Fig.~1,
where $x_1$ and $x_2$ are gate voltages
and $x_3$ is the magnetic flux through the loop.
The generalized forces are conventionally defined
as $F^k=-{\partial {\cal H}}/{\partial x_k}$.
Note that $F^3$ is the electrical current.
In LRT \cite{landau} the first order contribution to $\langle F^{k} \rangle$
is related to $x_j(t)$ by a causal
response kernel \mbox{$\alpha^{kj}(t-t')$}.
The Kubo expression for this response kernel is
$\alpha^{kj}(\tau) = \Theta(\tau) \ K^{kj}(\tau)$, where
$K^{kj}(\tau) = ({i}/{\hbar}) \langle [F^k(\tau),F^j(0)]\rangle$,
and $\Theta(\tau) $ is the step function.
The Fourier transform of $\alpha^{kj}(\tau)$
is the generalized susceptibility $\chi^{kj}(\omega)$.
The generalized conductance matrix is:
\begin{eqnarray} \label{e1}
\bm{G}^{kj} \ = \ \lim_{\omega\rightarrow 0}
\frac{\im[\chi^{kj}(\omega)]}{\omega} \ = \
\int_0^{\infty} K^{kj}(\tau)\tau d\tau
\end{eqnarray}
Thus in the limit of zero frequency the non-trivial
part of the response can be written as a generalized Ohm law
\begin{eqnarray} \label{e2}
\langle F^k \rangle = -\sum_j \bm{G}^{kj} \dot{x}_j
\ \equiv \ \left(-\bm{\eta} \cdot \dot{x} - \bm{B}\wedge \dot{x} \right)_{k}
\end{eqnarray}
where following \cite{robbins}
the generalized conductance matrix is
written as a sum of a symmetric
matrix $\bm{\eta}^{kj}=\bm{\eta}^{ik}$ that represents the
dissipative response, and an antisymmetric
matrix $\bm{B}^{kj}=-\bm{B}^{jk}$ that
represents the non-dissipative response
(also called ``geometric magnetism").

For a device as in Fig.1, and zero temperature occupation 
of non-interacting (spinless) Fermions, 
we find below that the dissipative part of the response is
\begin{eqnarray} \label{e3}
\bm{\eta}^{kj} &=&
\frac{\hbar}{\pi}\ \trc \left[F^k \ \im[{\mathsf G}^{+}] \  F^j \ \im[{\mathsf G}^{+}] \right]
\\ \label{e4}
&=& \frac{\hbar}{4\pi}
\trc \left[ \frac{\partial S^{\dag}}{\partial x_i} \frac{\partial S}{\partial x_j}\right]
\end{eqnarray}
where ${\mathsf G}^{\pm}=1/(E{-}{\cal H}{\pm}i0)$ are Green functions of 
the corresponding open system, and 
$\im [{\mathsf G^{+}}] = -i\half({\mathsf G}^{+}{-}{\mathsf G}^{-})$.
For the non-dissipative part of the response we find:
\begin{eqnarray} \label{e5}
\bm{B}^{kj} &=&
-\frac{i\hbar}{2\pi}\ \trc \left[F^k \ ({\mathsf G}^{+}{+}{\mathsf G}^{-}) \  F^j \ \im[{\mathsf G}^{+}] \right]
\\ \label{e6}
&=& \frac{e}{4\pi i}
\trc \left[ P_{\tbox{A}}\left(
\frac{\partial S}{\partial x_j} S^{\dag}
-\frac{\partial S^{\dag}}{\partial x_j} S
\right)\right]
+\bm{B}^{3j}_{\tbox{intrf}}
\end{eqnarray}
where the second equality holds for $k=3$,
and allows the determination of the electrical current $\langle F^3 \rangle$
via a specified lead~$\mbox{\footnotesize A}$.
The last term is defined in Eq.(\ref{e25}).  
The projector $P_{\tbox{A}}$ restricts the trace operation
to be over the specified lead channels.
In the absence of magnetic field the remaining
component is $\bm{B}^{12}=0$,
while $\bm{\eta}^{31}=\bm{\eta}^{32}=0$ as expected from
the Onsager reciprocity relations (see last paragraph).
Disregarding the last term in Eq.(\ref{e6}), 
the sum of (\ref{e6}) and (\ref{e4}) 
for $k=3$ coincides with the BPT formula,
which can be written in our notations as:
\begin{eqnarray} \label{e7}
\bm{G}^{3j} = \frac{e}{2\pi i}
\trc\left(P_{\tbox{A}}\frac{\partial S}{\partial x_j}
S^{\dag}\right)
\ \ \ \ \mbox{\small [BPT]}
\end{eqnarray}
We show later that this reduces for $j=3$
to the Landauer formula which relates the electrical
conductance $\bm{G}^{33}$ to the transmission of the device.


Below we explain how to derive the expressions
for $\bm{\eta}^{kj}$ and $\bm{B}^{kj}$ starting from the Kubo formula Eq.(\ref{e1}).
Later we discuss further physical implications of our results. 
Assuming zero temperature Fermi occupation
up to energy $E_F$, standard textbook procedure 
\cite{landau,imry,datta,imryK} leads to
\begin{eqnarray} \label{e8}
\bm{\eta}^{kj}\Big|_{\Gamma} = \pi\hbar \sum_{n,m}
F^k_{nm} \ \overline{\delta(E_F-E_m)} \ F^j_{mn} \ \overline{\delta(E_F-E_n)}
\end{eqnarray}
where the overline indicates that the delta functions are smeared. 
If the system were not isolated, the ``broadening" $\Gamma$ of the 
energy levels would be determined by the interaction with 
the external environment \cite{imryK}. 
But we assume a closed {\em isolated} system. 
Still we argue \cite{pmc} that in case 
of a quantized {\em chaotic} system the levels
acquire an effective width
\mbox{$\Gamma=(({\hbar\sigma_F}/{\Delta^2})|\dot{x}|)^{2/3}\Delta$},  
where $\Delta$ is the mean level spacing, 
and $\sigma_F$ is the root-mean-square value of 
the near-diagonal matrix elements (see remark \cite{note}).   
Therefore we find two possibilities: 
In the adiabatic regime ($\Gamma \ll \Delta$) the 
dissipative conductance is zero ($\bm{\eta}=0$), 
while in the non-adiabatic regime ($\Gamma > \Delta$) 
the dissipative conductance acquires a well defined 
finite value, which is {\em not} sensitive to $\Gamma$, 
and can be calculated using Eq.(\ref{e3}). 
A similar claim holds regarding $\bm{B}^{kj}$, 
but the details are much more subtle: 
We start with the standard expression \cite{robbins,pmp}
\begin{eqnarray} \label{e9}
\bm{B}^{kj}\Big|_{\Gamma{=}0} = 2\hbar \sum_n f(E_n) 
\sum_{m(\ne n)}
\frac{\im\left[
F^k_{nm}F^j_{mn}\right]}
{(E_m-E_n)^2}
\end{eqnarray}
where $f(E)$ is the Fermi occupation function 
(later we take the limit of zero temperature).  
Incorporating $\Gamma$, and exploiting 
the antisymmetry of the numerator with respect 
to $n\Leftrightarrow m$ interchange we get 
\begin{eqnarray} \label{e10}
\bm{B}^{kj}\Big|_{\Gamma}  =  \sum_{n,m}
\frac{ -i\hbar F^k_{nm}F^j_{mn}}
{(E_m{-}E_n)^2 {+} (\Gamma/2)^2}(f(E_n){-}f(E_m))
\end{eqnarray}
The numerator, on the average, depends mainly 
on the difference $r=m-n$, and it is non-negligible 
within a bandwidth $|E_m-E_n|<\Delta_b$. 
We further discuss the bandwidth issue
in the next paragraph, and explain  
that in the limit of a very long wire 
$\Delta \ll \Gamma \ll \Delta_b$. 
This means that in this limit $\Gamma$ 
serves like the infinitesimal $i0$ in the definition 
of the Green functions ${\mathsf G}^{\pm}$. 
Consequently, the sum in Eq.(\ref{e10}), 
which is of the form 
$\sum_{n,m}g(n{-}m)(f(E_n){-}f(E_m)) = \sum_r r g(r)$,   
leads after some straightforward algebra to Eq.(\ref{e5}).


Formally there is an optional derivation that leads to 
(\ref{e3}) and (\ref{e5}). The kernel $K^{ij}(\tau)$ 
is related to the symmetrized correlation function 
$C^{ij}(\tau) = \langle \half (F^i(\tau)F^j(0)+F^j(0)F^i(\tau)\rangle$. 
The quantum mechanical derivation of this subtle relation 
is discussed in Appendix~D of \cite{pmp}. If we use this relation 
we get from Eq.(\ref{e1}) an extremely simple (and useful) result: 
\begin{eqnarray} \label{eFD} 
\bm{G}^{kj} = \frac{1}{\Delta}\int_0^{\infty}C^{kj}(\tau)d\tau 
\end{eqnarray}
which can be regarded as the generalization 
of the {\em fluctuation dissipation relation}.   
The fluctuations are described by $\tilde{C}^{kj}(\omega)$ 
which is defined as the Fourier transform of $C^{ij}(\tau)$. 
It follows from this definition that 
\begin{eqnarray}
\tilde{C}^{kj}(\omega) = \frac{2\pi\hbar}{\Delta} 
\ \overline{F_{nm}^k F_{mn}^j}
\ \Big|_{E_n{-}E_m \approx \hbar\omega}
\end{eqnarray}
For the device of Fig.1 the mean level spacing is 
$\Delta \propto 1/L$ where $L$ is the length of the wire. 
The above relation implies that the bandwidth 
of the $mn$~matrix is $\Delta_b\sim\hbar/\tau_{cl}$, 
where the classical correlation time $\tau_{cl}$ 
is determined by the chaotic motion inside the dot. 
It is also clear that $\tilde{C}^{ij}(\omega) \propto 1/L$, 
and therefore $\sigma_F^2 \propto (1/L)^2$. 
Hence we get that $\Gamma \propto (1/L)^{1/3}$,  
implying that the limit $L\rightarrow\infty$ 
(keeping constant Fermi energy) is non-adiabatic, 
and that $\Delta\ll\Gamma\ll\Delta_b$. 
Assuming for simplicity that there is no magnetic field, 
one easily derives the expressions
\begin{eqnarray}
\bm{G}^{33} &=& \frac{1}{2\Delta} \ \tilde{C}^{33}(\omega \sim 0) \\ 
\bm{G}^{3j} &=& \frac{1}{\Delta} \int_{-\infty}^{\infty}
\Im\left[\frac{\tilde{C}^{3j}(\omega)}{\omega}\right] 
\frac{d\omega}{2\pi} \hspace*{0.7cm} \mbox{for $j{=}1,2$} 
\end{eqnarray}
which are equivalent to those obtained 
in the previous paragraph. Note that $C^{3j}(\tau)$ 
with $j=1,2$ is antisymmetric with respect to $\tau$, 
and therefore $-i\tilde{C}^{3j}(\omega)/\omega$ 
is a real symmetric function.


We turn back to the formal derivation. We want to get 
exact expressions for the elements of the 
conductance matrix, for the device of Fig.~1, 
in the non-adiabatic limit of large $L$.
The location of the particle is specified by 
\mbox{$\bm{r}=(\ar,s)$},
where $\ar$ is the coordinate along the ring,
and $s$ is a transverse coordinate.
Optionally we can specify the location along a lead using
a radial coordinate $r$, while the surface coordinate $s$
distinguishes different points that have the same $r$.
We shall refer to $r=0$ as the boundary of the scattering region.
The channel basis is defined as
$\langle \ar, s|a, r \rangle  =  \chi_a(s) \ \delta(\ar-\ar_a(r))$,
where $a$ is the channel index.
The wavefunction in the lead regions
can be expanded as follows:
\begin{eqnarray} \label{e14}
|\Psi\rangle = \sum_{a,r} \left(
C_{a,+} \eexp{ik_a r} + C_{a,-} \eexp{-ik_a r} \right)
\  |a, r\rangle
\end{eqnarray}
Following \cite{datta} we define an operator
which can be  identified with the imaginary part of the self energy
of the interaction of the dot with the leads:
\begin{eqnarray} \label{e15}
\hat{\Gamma} = \sum_a |a, 0\rangle \hbar v_a  \langle a, 0|
= \delta(r) \otimes \sum_a |a\rangle \hbar v_a  \langle a|
\end{eqnarray}
where $v_a=(\hbar k_a/\mbox{mass})$ is the velocity in channel ~$a$.
The matrix elements of the second term in Eq.(\ref{e15}) are
\begin{eqnarray} \label{e16}
\hat{\Gamma}(s,s') =   \sum_a  \chi_a(s) \ \hbar v_a \ \chi_a(s')
\end{eqnarray}
Using standard procedure (see section~3.4 in \cite{datta})
the Green function in the leads, inside the scattering region ($r<0$),
can be expressed using the $S$~matrix:
\begin{eqnarray} \label{e17}
{\mathsf G}^{+}(r,s|0,s') = \hspace*{0.5\hsize}
\\ \nonumber
- i\sum_{a,b}
\chi_b(s) \frac{1}{\sqrt{\hbar v_b}} (\eexp{-ikr}+S\eexp{ikr})_{ba}
\frac{1}{\sqrt{\hbar v_a}} \chi_a(s')
\end{eqnarray}
where $k=\mbox{diag}\{k_a\}$ is a diagonal matrix.
Now we are fully equipped to convert Eq.(\ref{e3}) 
into an $S$-matrix expression.
Using the  identities (see \cite{datta} for Eq.(\ref{e18}))
\begin{eqnarray} \label{e18}
\im[{\mathsf G}^{+}]
&=& -\half {\mathsf G}^{+} \hat{\Gamma} {\mathsf G}^{-}
= -\half {\mathsf G}^{-} \hat{\Gamma} {\mathsf G}^{+}
\\ \label{e19}
\frac{\partial {\mathsf G}^{\pm}}{\partial x_j} &=& - {\mathsf G}^{\pm}F^j {\mathsf G}^{\pm}
\end{eqnarray}
we obtain 
\begin{eqnarray} \label{e20}
\bm{\eta}^{kj} = \frac{\hbar}{4\pi} \trc\left[
\frac{\partial {\mathsf G}^{+}}{\partial x_j} \hat{\Gamma}
\frac{\partial {\mathsf G}^{-}}{\partial x_j}  \hat{\Gamma}
\right]
\end{eqnarray}
Using the definition of $\hat{\Gamma}$ and Eq.(\ref{e17})
we get Eq.(\ref{e4}).

The derivation of the $S$ matrix expression Eq.(\ref{e6}) for $\bm{B}^{kj}$
is much more subtle, and requires a preliminary discussion
of the definition of the current operator.
Consider a ring geometry, and assume that the current is driven
by the flux $\Phi$. In order to have a better defined model
we should specify what is the vector potential ${\cal A}(\bm{r})$ along the ring.
We can regard the values of ${\cal A}$ at different points in space
as independent parameters (think of tight binding model).
Their sum (meaning $\oint {\cal A}(\bm{r}) {\cdot} d\bm{r}$) should be $\Phi$.
So we have to know how $\Phi$ is ``distributed" along the ring.
This is not just a matter of ``gauge choice" because
the electric field \mbox{${\cal E}(\bm{r}) = - \dot{{\cal A}}(\bm{r})$}
is a gauge invariant quantity.
The transformation ${\cal A} \mapsto {\cal A}+\nabla\Lambda(\bm{r})$
for a time dependent filed is not merely a gauge change:
A gauge transformation of time dependent field requires
a  compensating replacement of the scalar potential, which is not the case here.
So let us define a flux $\Phi_A$ which is associated with
a vector potential that is concentrated across
a section $\ar=\ar_A$ of a given lead.
For the later derivation it is essential to assume that the
section $\ar=\ar_A$ is contained within the scattering region (see Fig.~1).
The generalized force which is associated with $\Phi_A$
is $F^3=I_A$, the current through this section. Namely
\begin{eqnarray} \label{e21}
I_A = -\frac{\partial {\cal H}}{\partial \Phi_A}  &=&
\half e (v \ \delta(\ar-\ar_A) + \delta(\ar-\ar_A) v)
\\ \label{e22}
&=& (e/\hbar)[ \hat{\Gamma}_A P^{+} - \hat{\Gamma}_A P^{-} ]
\end{eqnarray}
where $v$ is the $\ar$ component of the velocity operator.
The last equality involves new definitions. We define
\begin{eqnarray} \label{e23}
\hat{\Gamma}_A = \sum_{a \in A} |a, r_A\rangle \hbar v_a  \langle a, r_A|
\end{eqnarray}
We also define projectors $P^{+}$ and $P^{-}$ that project
out of the lead wavefunction Eq.(\ref{e14}) the outgoing and the ingoing
parts respectively. These projectors commute with $\hat{\Gamma}_{\tbox{A}}$.
Furthermore, note that  $P^{+}{\mathsf G}^{+} = {\mathsf G}^{+} $,
and that $P^{-}{\mathsf G}^{+} = 0 $,
and that ${\mathsf G}^{-}P^{-} = 0 $ etc.
Using these extra identities one obtains the following expression:
\begin{eqnarray} \label{e24}
\bm{B}^{3j} = \frac{e}{4\pi i} \trc\left[
\hat{\Gamma}_{\tbox{A}}\frac{\partial {\mathsf G}^{+}}{\partial x_j}
\hat{\Gamma} {\mathsf G}^{-}
-\hat{\Gamma}_{\tbox{A}} \frac{ \partial {\mathsf G}^{-}}{\partial x_j}
\hat{\Gamma} {\mathsf G}^{+}
\right]
\end{eqnarray}
Using the definitions of $\hat{\Gamma}$ and $\hat{\Gamma}_{\tbox{A}}$,
together with Eq.(\ref{e17}), followed by a straightforward
algebraic manipulation, one arrives at Eq.(\ref{e6}) with the
additional term
\begin{eqnarray} \label{e25}
\bm{B}^{3j}_{\tbox{intrf}} =
\frac{e}{2\pi} \re\left[
\trc \left(
P_{\tbox{A}} \frac{\partial S}{\partial x_j}\eexp{i2kr_{\tbox{A}}}
\right)\right]
\end{eqnarray}
where $P_{\tbox{A}}$ is a projector that restrict the trace
operation to the $a\in A$ lead channels.

For the simple ring geometry of Fig.~1,
we have a left lead ($b\in B$) and a right lead ($a\in A$)
channels, and the $S$-matrix can be written as
\begin{eqnarray} \label{e26}
S = \left( \matrix{
\bm{r}_{\tbox{B}}  & \bm{t}_{\tbox{AB}}\eexp{-i\phi} \cr
\bm{t}_{\tbox{BA}} \eexp{i\phi} & \bm{r}_{\tbox{A}} }\right),
\ \ \ \ \ \ \
P_{\tbox{A}} = \left( \matrix{\bm{0} & \bm{0} \cr \bm{0} & \bm{1}} \right)
\end{eqnarray}
where $\phi=e\Phi_{\tbox{A}}/\hbar$. Using the identity
\begin{eqnarray} \label{e27}
\frac{\partial S}{\partial \Phi_{\tbox{A}}} \ = \  i\frac{e}{\hbar}(P_{\tbox{A}}S-SP_{\tbox{A}})
\end{eqnarray}
one can derive the relation that has been stated
between Eq.(\ref{e4}), Eq.(\ref{e6}) and the BPT formula Eq.(\ref{e7}).
Furthermore, assuming that there is an electro-motive force $-\dot{\Phi}_B$
which is induced in the other lead, one obtains from BPT
$\bm{G}^{33} = ({e^2}/{2\pi\hbar})\trc(\bm{t}_{\tbox{AB}}\bm{t}_{\tbox{AB}}^{\dag})$
which is the Landauer formula. The application of this procedure
to multi-lead systems is a straightforward generalization.

For an {\em open} system it is evident that the current which is emitted (say) through the right
lead, does not have to be equal to the current which is absorbed by the left lead.
The reason is that charge can be accumulated in the dot region.
But for a pumping cycle one realizes that the integrated current (pumped charge)
is a well defined (lead independent) quantity. Similar observation holds
in case of a {\em closed} system. Assume for example that the left lead is blocked.
In such case raising the dot potential will cause an emission of charge through
the right lead, while the current through the left lead is zero. The emitted charge
is accumulated in the ``wire". But for a full cycle the original charge distribution is restored,
and therefore the integrated charge ($Q$) becomes a well defined (lead independent) quantity.
The additional term Eq.(\ref{e25}) gives a zero net contribution for a full
pumping cycle. This term implies that the current is not uniform within the lead.
The current has a modulation in the radial direction ($r$), with a spatial period
that equals half the De-Broglie wavelength at the Fermi energy.
This reflects that the net transported current corresponds
to translation of a standing wave which is associated with the last occupied level.

More subtle is the value of $Q$ for a full driving cycle.
In contrast to a previous wrong statement \cite{SAA}
we have argued \cite{pmp} that for a strictly adiabatic driving
cycle, in the absence of magnetic field,  
the transported charge $Q$ is at best {\em approximately} quantized 
(say $Q\approx 1$ in units of the elementary charge).
The deviation is related to the Thouless conductance of the device,  
and can be either positive or negative \cite{pmp}. In contrast to that,   
with the BPT formula the correction to $Q\approx 1$ is always negative.  
On the basis of our derivation we can conclude 
the following: The deviation from quantization in a strictly 
adiabatic cycle is related to the contribution of the neighboring level. 
If the degeneracy with this level is located in the 
plane $(x_1,x_2,x_3{=}0)$ of the encircling cycle,  
then the correction is positive. If the encircled degeneracy is off plane,  
then the correction is negative.  The effect of non-adiabaticity 
($\Gamma > \Delta$) is to screen the contribution of the neighboring 
levels, which is the reason for having always a negative correction 
from the BPT formula.

The role that dissipation may have in pumping is restricted,
merely by the realization that the BPT formula is related to the Kubo formula.
The Onsager reciprocity relation imply that in the absence 
of magnetic field the conductance matrix $\bm{G}^{kj}$ should
be symmetric (antisymmetric) with respect to
the permutation of the indexes $(k,j)$,
depending on whether $F^k$ and $F^j$ transform
(not) in the same way under time reversal.
This means that shape deformations lead to dissipation 
via $\bm{\eta}^{ij}$ with $i{,}j \ {<} 3$, while the electrical current 
is determined exclusively by the non-dissipative 
terms $\bm{B}^{31}$ and $\bm{B}^{32}$.
This should be contrasted with the response to
electro-motive force which is purely 
dissipative: Both the current and the dissipation are 
exclusively determined by the Ohmic conductance $\bm{\eta}^{33}$.
Thus, in the absence of magnetic field, we have a clear
cut distinction between the dissipative and the non-dissipative 
contributions to the response.

In summary, starting with the Kubo formalism,
we were able to find expressions for the dissipative
and for the non-dissipative parts of the response,
and to illuminate the role of non-adiabaticity in the limiting case
of an {\em infinite} system. In contradiction with past
speculations, we were able to demonstrate that the switch
to an {\em open} system does not necessitate an extra dissipative term.

It is my pleasure to thank T.~ Dittrich (Colombia), T.~ Kottos (Gottingen),
and H.~Schantz (Gottingen) for useful discussions. This research
was supported by the Israel Science Foundation (grant No.11/02),
and by a grant from the GIF, the German-Israeli Foundation for Scientific
Research and Development.



\vspace*{1cm}

\centerline{\epsfig{figure=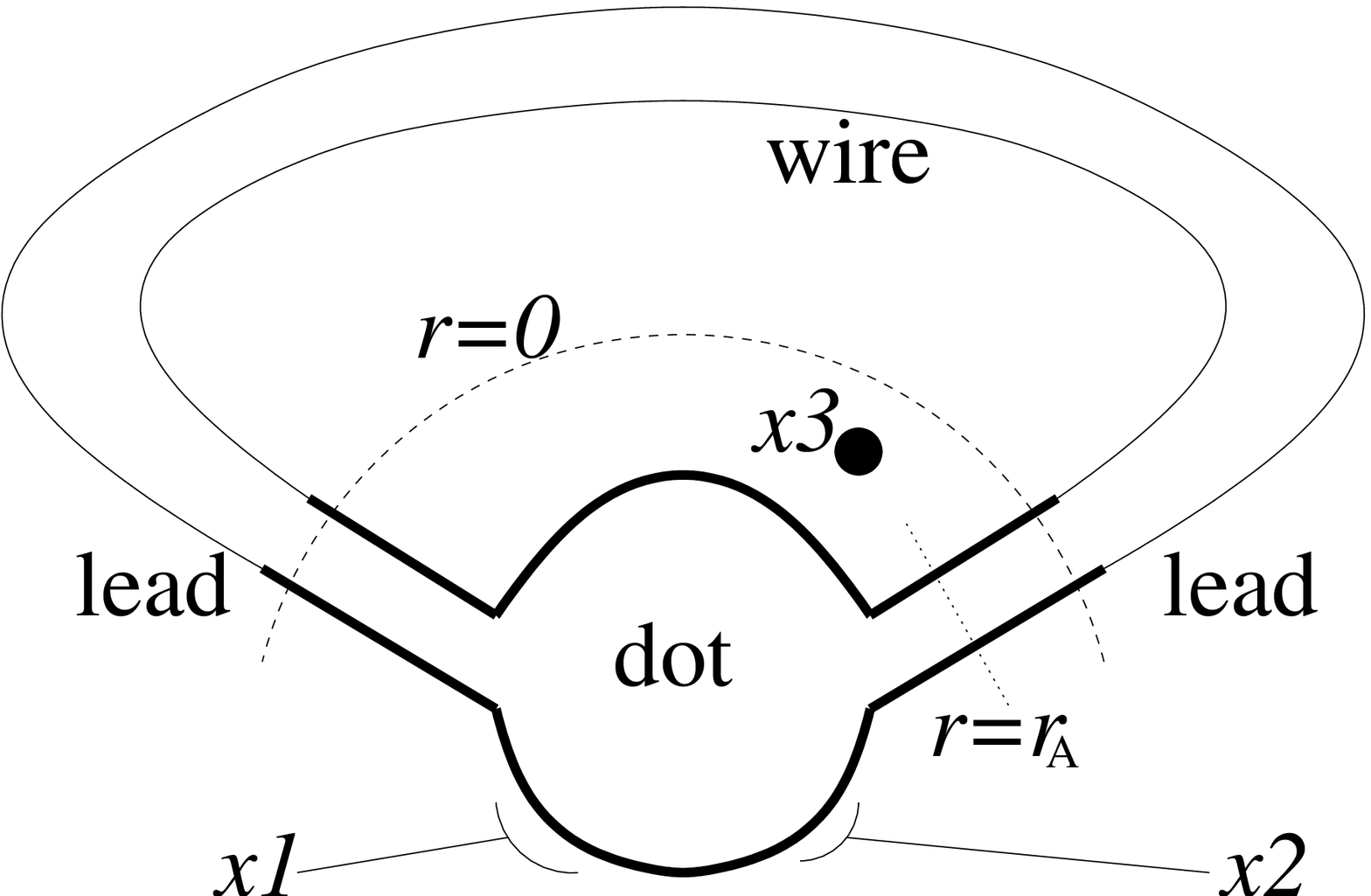,width=0.7\hsize}}
{\footnotesize FIG1.
Illustration of a closed system.
The dot potential is controlled
by gate voltages $x_1$ and $x_2$.
The flux through the loop is $x_3{=}\Phi$.
The scattering region ($r{<}0$)
is represented by an $S$~matrix. 
The length ($L$) of the wire is assumed 
to be very large.}

\clearpage
\onecolumngrid

{\LARGE NOT PART OF THE PAPER}

\ \\ \ \\ \ \\ 

\appendix 

\Cn{\Large\bf Going from Eq.(10) to Eq.(5)}

\ \\
In this appendix we give 
the ``straightforward algebra" 
that leads from Eq.(10) to Eq.(5).
\begin{eqnarray} \nonumber
\bm{B}^{kj} \ = \  -i\hbar\sum_{n,m}
\frac{F^k_{nm}F^j_{mn}}
{(E_m{-}E_n)^2 {+} (\Gamma/2)^2}(f(E_n){-}f(E_m))
\end{eqnarray}
Within the framework of the approximation which is discussed 
in the text this sum is of the form 
\begin{eqnarray} \nonumber
\bm{B}^{kj} \ = \ \sum_{n,m}
g(n-m) \times (f(E_n){-}f(E_m))
\end{eqnarray}
Using the notation $r=m-n$ it leads to  
\begin{eqnarray} \nonumber
\bm{B}^{kj}  \ = \ \sum_{r} 
g(r) \times r
\end{eqnarray}
We make the substitutions 
\begin{eqnarray} \nonumber
r &\ = \ & \frac{E_{m}-E_n}{\Delta} 
\\ \nonumber
g(r) &\ = \ & \frac{ -i\hbar F^k_{nm}F^j_{mn}}
{(E_m{-}E_n)^2 {+} (\Gamma/2)^2}
\end{eqnarray}
where $n$ is the index of 
an arbitrary energy level 
in the vicinity of the Fermi energy 
($E_n\sim E_F$).  \\ 
Now we have the expression
\begin{eqnarray} \nonumber
\bm{B}^{kj} \ = \ -i\hbar \frac{1}{\Delta}\sum_{m}
F^k_{nm}
\frac{E_m-E_n}
{(E_m{-}E_n)^2 {+} (\Gamma/2)^2}
F^j_{mn}
\end{eqnarray}
Thanks to $\Gamma$ we can make the replacement $E_n \mapsto E_F$. \\
Next we can use (in reverse) the approximation 
\begin{eqnarray} \nonumber
\sum_n \langle n |A|n \rangle \overline{\delta(E_n-E_F)} \ \approx \  
\frac{1}{\Delta}\langle n |A|n \rangle  
\end{eqnarray}
where the overline indicates a smeared delta function. \\
Hence we get
\begin{eqnarray} \nonumber
\bm{B}^{kj} \ = \  -i\hbar \sum_{m}
F^k_{nm}
\frac{E_m-E_F}
{(E_m{-}E_F)^2 {+} (\Gamma/2)^2}
F^j_{mn}
\ \overline{\delta(E_n-E_F)}
\end{eqnarray}
As explained in the text, in the limit 
of a very long wire $\Gamma$ is like 
the infinitesimal $i0$. \\
Hence we get Eq.(5).

\clearpage
\onecolumngrid

{\LARGE APPENDIX D OF REF.[8]}

\ \\ \ \\ \ \\ 

\Cn{\Large\bf Expressing $\tilde{K}(\omega)$ using  $\tilde{C}(\omega)$}

We can use the following manipulation in order
to relate  $\tilde{K}^{ij}(\omega)$ to $\tilde{C}^{ij}(\omega)$,
\begin{eqnarray} \label{eC1}
\tilde{K}^{ij}(\omega)&=& \sum_n f(E_n) \ \tilde{K}^{ij}_n(\omega)
\\ \nonumber &=&
\frac{i}{\hbar} 2\pi \sum_{nm}f(E_n)
(F^i_{nm}F^j_{mn}\delta(\omega+\omega_{nm})-F^j_{nm}F^i_{mn}\delta(\omega-\omega_{nm}))
\\ \nonumber  &=&
\frac{i}{\hbar} 2\pi \sum_{nm}f(E_m)
(-F^i_{nm}F^j_{mn}\delta(\omega+\omega_{nm})+F^j_{nm}F^i_{mn}\delta(\omega-\omega_{nm}))
\\ \nonumber  &=&
\frac{i}{\hbar} 2\pi \sum_{nm} \frac{f(E_n)-f(E_m)}{2}
(F^i_{nm}F^j_{mn}\delta(\omega+\omega_{nm})-F^j_{nm}F^i_{mn}\delta(\omega-\omega_{nm}))
\\ \nonumber  &=&
-i \omega \pi \sum_{nm}\frac{f(E_n)-f(E_m)}{E_n-E_m}
(F^i_{nm}F^j_{mn}\delta(\omega+\omega_{nm})+F^j_{nm}F^i_{mn}\delta(\omega-\omega_{nm}))
\\ \nonumber &=&
-i\omega\sum_{n} f'(E_n) \ C^{ij}_n(\omega)
\end{eqnarray}
where we use the notation $\omega_{nm}=(E_n-E_m)/\hbar$.
The third line differs from the second line by permutation of the dummy
summation indexes, while the fourth line is the sum of the second
and the third lines divided by 2. In the last equality we assume small $\omega$.
If the levels are very dense, then we can replace the summation by integration,
leading to the relation:
\begin{eqnarray} \label{eKC}
\tilde{K}^{ij}(\omega) \ \ \equiv \ \ 
\int g(E)dE \ f(E) \ \tilde{K}^{ij}_{E}(\omega)
\ \ = \ \
-i \omega \int g(E)dE \ f'(E) \ \tilde{C}^{ij}_{E}(\omega)
\end{eqnarray}
where $\tilde{K}^{ij}_{E}(\omega)$ and $\tilde{C}^{ij}_{E}(\omega)$
are microcanonically smoothed functions.
Since this equality hold for any smooth $f(E)$, it follows
that the following relation holds (in the limit $\omega\rightarrow0$):
\begin{eqnarray}
\tilde{K}^{ij}_E(\omega)
\  = \
i\omega \frac{1}{g(E)}\frac{d}{dE}\left[g(E)C^{ij}_E(\omega)\right]
\end{eqnarray}
If we do not assume small $\omega$, but instead assume canonical state,
then a variation on the last steps in Eq.(\ref{eC1}),
using the fact that $(f(E_n){-}f(E_m))/ (f(E_n){+}f(E_m))= \tanh((E_n{-}E_m)/(2T))$
is an odd function, leads to the relation
\begin{eqnarray}
\tilde{K}^{ij}_T(\omega)  \ = \  i\omega \times
\frac{1}{\hbar\omega}\tanh\left(\frac{\hbar\omega}{2T}\right)  \  C^{ij}_T(\omega)
\end{eqnarray}
Upon substitution of the above expressions in the Kubo formula for $\bm{G}^{ij}$,
one obtains the Fluctuation-Dissipation relation.

\ \\ \ \\

{\bf Note added:} For a low temperature Fermi occupation Eq.(\ref{eKC}) can be written as  
\begin{eqnarray} 
\tilde{K}^{ij}(\omega) \ \ = \ \
i \omega \ g(E_F) \ \tilde{C}^{ij}_{E_F}(\omega)
\end{eqnarray}
Formally this is valid only if $\hbar\omega \ll T$. But in fact 
if $T\ll E_F$ the result should not be sensitive to $E_F$.  
Therefore it can be argued that this relation holds globally, 
and we can set $T=0$. It follows that 
\begin{eqnarray} 
K^{ij}(\tau) \ \ = \ \
-\frac{\partial}{\partial \tau} \ g(E_F) \ \tilde{C}^{ij}_{E_F}(\tau)
\end{eqnarray}
This can be used in order to derive Eq.(11) for Eq.(1). 
We see that it is a subtle relation which should not be regarded as 
a trivial identity.


\begin{thebibliography}{99}
\bibitem{note}
Any segment of a pumping cycle in the $(x_1,x_2)$ plane 
can be regarded as ``one parameter driving" with $x$ 
a linear combination of $x_1, x_2$, while $F$ is its conjugate.
\end{thebibliography}
\end{document}